\documentclass[preprint2]{aastex}
\usepackage{psfig,natbib}
\shortauthors{Bower et al.}
\shorttitle{Turbulent Accretion onto Sagittarius A*}
\begin{document}

\newcommand\degd{\ifmmode^{\circ}\!\!\!.\,\else$^{\circ}\!\!\!.\,$\fi}
\newcommand{\etal}{{\it et al.\ }}
\newcommand{\uv}{(u,v)}
\newcommand{\rdm}{{\rm\ rad\ m^{-2}}}
\newcommand{\msuny}{{\rm\ M_{\sun}\ y^{-1}}}
\newcommand{\mylesssim}{\stackrel{\scriptstyle <}{\scriptstyle \sim}}
\newcommand{\sci}{Science}


\title{Variable Linear Polarization from Sagittarius A*:  Evidence for
a Hot Turbulent Accretion Flow}

\author{Geoffrey C. Bower\altaffilmark{1}, 
Heino Falcke\altaffilmark{2},
Melvyn C.H. Wright\altaffilmark{1},
Donald C. Backer\altaffilmark{1}}

\altaffiltext{1}{Astronomy Department \& Radio Astronomy Laboratory, 
University of California, Berkeley, CA 94720; gbower,dbacker,wright@astro.berkeley.edu} 
\altaffiltext{2}{Radio Observatory Westerbork ,  ASTRON , P.O. Box 2 , 7990 AA Dwingeloo, The Netherlands; Visiting Scientist, Max-Planck-Institut f'r Radioastronomie, Auf dem H\"ugel 69, 53121 Bonn, Germany; Adjunct Professor, University of Nijmegen; falcke@astron.nl} 

\begin{abstract}
We report the discovery of variability in the
 linear polarization from the Galactic Center black hole source, 
Sagittarius A*.
New polarimetry obtained with the Berkeley-Illinois-Maryland Association
array at a wavelength of 1.3 mm shows a position angle
that differs by $28 \pm 5$ degrees from observations 6 months prior and
then remains stable for 15 months.  
This difference may be due to a change in the source emission region
on a scale of $10$ Schwarzschild radii or due to a change of $3\times 10^5\rdm$
 in the rotation measure.  We consider a change in the source physics
unlikely, however, since we see no corresponding change in the total intensity 
or polarized intensity fraction.    On the other hand,
turbulence in the accretion region at a radius $\sim 10$ to $1000  R_s$ 
could readily
account for the magnitude and time scale of the position angle change.
\end{abstract}


\section{Introduction}

The extreme underluminosity ($\sim 10^{-10}$ times the Eddington luminosity) 
of Sagittarius A*, 
the $3\times 10^6 M_\sun$ 
black hole in the Galactic Center, is a fundamental puzzle which has inspired many
theoretical efforts \citep[e.g.,][and references therein]{2001ARA&A..39..309M}.  Broadly, these can
be classified as low accretion rate models and low radiative efficiency
models.  The recent discovery of linear polarization at 
wavelengths of 1.3 mm and shorter \citep{2000ApJ...534L.173A,2003ApJ...588..331B}
has demonstrated that there is a very low accretion
rate $\sim 10^{-7} \msuny$ and that the underluminosity is not solely 
due to radiatively inefficient accretion.  However, the
density of gas at the Bondi radius suggests that the accretion rate
at large radii 
is higher by several orders of magnitude \citep{1999ApJ...517L.101Q}.  This problem 
is resolved theoretically by the presence of convection, a wind, or an
outflow that carries away much of the infalling
material before it reaches the black hole 
\citep{2002ApJ...573..749B,2003ApJ...592..767P,2003ApJ...592.1042I}.
The resulting accretion appears to be turbulent rather than smooth,
potentially leading to flux density variations \citep{2004Goldston}.

Millimeter wavelength
linear polarimetry has the power to probe the structure and
turbulent nature of the accretion medium.  Recent measurements with
the Very Long Baseline Array have shown that emission at millimeter wavelengths
originates very close to the black hole at a radius of 
$\sim 10$ Schwarzschild radii ($R_s$) 
\citep{2004Sci...304..704}.
The source of the millimeter and submillimeter 
emission is either the base of a jet \citep{2002A&A...383..854Y}
or the inner edge of a hot accretion disk \citep{2001ApJ...561L..77L}.
This emission, which is linearly polarized,
must
then propagate through the magnetized accretion region.
The position angle of linear polarization will undergo Faraday
rotation in the accretion region.  A sufficiently large rotation measure (RM)
will cause the linear polarization to disappear when averaged over a
given bandwidth.  The presence of linear
polarization at 1.3 mm, therefore, indicates a strong
upper limit on the RM.  
This upper limit on the RM provides a constraint on
the accretion rate, which is dependent on the radial structure of the
magnetic field and density.  For most models, however, the range of acceptable
accretion rates is on the order of $10^{-7} \msuny$ \citep{2000ApJ...545..842Q,2002A&A...388.1106B}.
Variations in the accretion rate or in the structure of the magnetic field
or particle density, potentially as a result of turbulence,
will change the RM, leading to a change in position angle with time.

We present here new linear polarimetry of Sgr A* obtained with
the Berkeley-Illinois-Maryland Association (BIMA) array at 1.3 mm.  In
previous observations, we found that the position angle remained
constant at $139 \pm 4$ degrees in four observations in March through
May 2002.  This position angle was $\sim 50$ degrees greater than
the position angle found with JCMT observations.
Our new observations show that the position angle decreased by $\sim 30$
degrees in the 6 months following the BIMA observations
and then remained relatively stable over 15
months.  We describe our observations and analysis in \S 2 and our
linear polarization results in \S 3.  We discuss the results and
give our conclusions in \S 4.

\section{Observations and Data Analysis}

Polarimetric observations of Sgr A* were obtained on five separate
dates at two frequency settings, one centered at 216 GHz and
the other at 230 GHz (Table~1).  The BIMA array was in C configuration
for observations in October 2002 and May 2003 producing a resolution for Sgr A* of approximately
$7 \times 3$ arcsec.  The array was in B
configuration for observations in January 2004 producing a resolution of
$3 \times 1$ arcsec.
The observations were performed in
a polarization switching mode that gives a full, calibrated measurement
of the four Stokes parameters in five minutes \citep{1999ApJ...527..851B}.
Observations were 4 to 5 hours in duration centered on transit, placing  Sgr A*
at a typical elevation of 15 to 21 degrees.

Polarization leakage solutions were obtained from
observations of the source 3C 279 at 230 GHz on 13
October 2002 and at 216 GHz on 18 October 2002.
The 216 GHz leakage terms are $\sim 10\%$, which is larger
than the 230 GHz leakage terms because the
quarter-wavelength polarizing grids are optimized for 230 GHz.
We determined the position angle of linear polarization for 3C279 to be
$27 \pm 1$ degrees at 216 GHz and $36 \pm 1$ at 230 GHz.
We found that our results for Sgr A*
did not vary significantly when we forced the linear polarization for 3C 279
at 216 and 230 GHz to be equal in the process of solving for the
polarization leakage terms.  Time variations in leakage terms introduce
no more than 1\% error in the polarization, which corresponds to an
error of 6 degrees in the position angle for a source that is 10\% 
polarized \citep{2003ApJ...588..331B}.
The leakage terms determined at 230 GHz
are similar to those determined previously on 28 February 2002, which
were used in \citet{2003ApJ...588..331B}.  We also found that our results for Sgr A* 
at 230 GHz did not depend on whether we used the 13 October 2002 or
28 February 2002 leakage terms.  For instance, the position angle 
of the lower sideband (215 GHz) in the 19 May 2003 experiment is identical at
$104 \pm 5$ degrees using either set of leakage terms.  We conclude that errors in
the leakage terms do not significantly alter our position angle at 
a level of 10 degrees.

The Sgr A* data were phase self-calibrated and averaged over 5 minute
intervals.  The appropriate leakage corrections were applied.
The flux density in each Stokes parameter was determined from
fits in the $\uv$ plane to data on baselines longer than 20 $k\lambda$.
For the B array experiments,
we found similar results using only baselines longer than 
40 $k\lambda$.  This indicates that our results are not corrupted by
polarized dust, confirming measurements and arguments previously
published \citep{2003ApJ...588..331B}.  We also compared 
results for the first and second-half of each experiment and found
no evidence for variability.  In addition, we found no dependence
on self-calibration interval.

Results are listed in Table~1.  We
give the best-fit value for each Stokes parameter in each sideband
and for the average of the two sidebands.  The fractional polarization
and position angle are calculated from Stokes $Q$ and $U$
for each sideband and for the average.

There is an apparent detection of circular polarization in the mean
of all experiments 
$-3 \pm 1\%$.  This may result from the failure of the linear 
approximation for polarization leakage which leads to terms for 
the circular polarization proportional to $D P$ and $D^2 I$, where
$D$ is a typical leakage term, $P$ is the linear polarization
and $I$ is the total intensity.  For $D \sim P \sim 10\%$, these
terms contribute a false circular polarization $\sim 1\%$, which is
comparable to the measured circular polarization.  Gain variations
may also contribute to a false circular polarization signal.

The range of total intensity flux densities 
from this paper and from our previous paper is  0.7 to 2.4 Jy,
which 
falls below the 1 to 4 Jy range measured by 
\cite{2003ApJ...586L..29Z}
at the same
frequency with the Submillimeter Array.  The mean in the BIMA
data is less by a factor $\sim 2$.  The origin of these
differences is uncertain but may be partly due to atmospheric 
phase decorrelation at
the BIMA site.  These variations in the flux density will not have
an effect on the polarization fraction or position angle because all
Stokes parameters are equally affected by the decorrelation.

\section{Linear Polarization Results}

Sgr A* is clearly detected in linear polarization in all epochs.  We show
the fractional polarization and the position angle as a function of time in
Figures~\ref{fig:pfrac} and \ref{fig:pangle}.  We also plot the results from
\citet{2000ApJ...534L.173A} and \citet{2003ApJ...588..331B}
in these Figures.  

The fractional polarization is apparently constant with time.  The mean
polarization fraction determined from these observations is $9.9 \pm 1.4\%$.
This is consistent within $2\sigma$ of the mean determined from our
previous observations of $7.2 \pm 0.6\%$.  The mean of all BIMA observations
is $7.5 \pm 0.5\%$.  If we exclude the last observation from \citet{2003ApJ...588..331B}
which appears to be an outlier, then the mean polarization fraction is $8.9 \pm 0.6\%$.  

The position angle is not constant with time.  The mean position angle from
these new observations is $111 \pm 3$ deg.  This differs sharply from the
mean of our past observations (March through May 2002) of $139 \pm 4$ deg, as well as from the Aitken
\etal 2000 value of $88 \pm 3$ deg (August 1999).  

We find estimates of the RM using the contemporaneous observations covering
215 to 230 GHz.  We perform a least squares fit to the position angle as a function of $\lambda^2$ for the four frequencies.  For the 14 and 17 October 2002 results, we find an RM
$0.8 \pm 1.6 \times 10^6  \rdm$.  For the 27 December 2003 and
5 January 2004 results, we find an RM $2.9 \pm 0.9 \times 10^6 \rdm$.
Averaging the Stokes parameters over time, we compute a
mean RM that is significant at the $3\sigma$ level: $2.4 \pm 0.8
\rdm$.  If we include a 10 degree systematic error in the position angle
at 216 GHz, however, the significance of this result drops to $\sim 1\sigma$.
Considering the possibility of additional error originating from 
the polarization leakage terms, we consider this estimated RM to be
an upper limit rather than a detection.

The upper limit to the RM $\sim 2 \times 10^6 \rdm$
is consistent with results determined from previous
observations.  Relative to previous BIMA observations, these results set an upper limit based on a broader frequency range, 215 to 230 GHz as opposed to 227
to 230 GHz \citep{2003ApJ...588..331B}.  The absence of significant bandpass depolarization in JCMT measurements produces a comparable result 
\citep{2000ApJ...534L.173A}.  
Together, these results support the conclusion that the mass accretion rate onto Sgr A* has an upper limit of $\sim 10^{-7} M_{\sun} {\rm\ y^{-1}}$,
eliminating ADAF and Bondi-Hoyle models.

The variability in the polarization angle 
invalidates previous
determinations of source physics based on non-contemporaneous
measurements.  These include the apparent $\sim 90$ degree position angle 
jump in the JCMT results \citep{2000ApJ...534L.173A} 
as well as RM estimates \citep{2003ApJ...588..331B}.
In fact, even our estimated RM from observations
separated by 1 week must be viewed as potentially corrupted by
variability.  These results do suggest that the difference between
the JCMT position angle at 220 GHz and previous BIMA results at
230 GHz is due to variability.

\section{Discussion}

There are two possible interpretations for the time variability of
the position angle:  a change in the polarization of the source; or, 
a change in the medium through which the polarization propagates.  In 
the intrinsic polarization scenario, the magnetic field structure 
from which the polarized radiation originates must undergo a
change.  This might be due to the propagation of
shock in the jet or a change in the orientation of a thin disk.
Variability in the centimeter wavelength circular polarization has been
interpreted as the result of intrinsic source variations 
\citep{2002A&A...388.1106B}.
In the propagation scenario, turbulence or clumpiness
in the accretion region can change the RM, which then alters
the position angle of the linear polarization.   The necessary change
in the RM is $\Delta {\rm RM}\sim 3 \times 10^5 \rdm$.  

Although
both scenarios are possible, the apparent stability of the fractional
polarization leads us to favor a changing RM as the explanation.  
Typically, a changing polarization position angle in a jet from a 
shock is accompanied by a sharp change in the total intensity and polarization
fraction \citep{1985ApJ...298..114M}.  In a disk model for the origin
of the linear polarization the polarization fraction  in the optically thin
limit is highly variable
on a time scale of days to weeks while the polarization vector is
quite stable \citep{2004Goldston}.  

Synchrotron self-absorption has also been proposed as the source of wavelength-dependent change in the position angle 
\citep{2000ApJ...534L.173A,2000ApJ...538..L121}.  
A change in the self-absorption frequency would lead to a change in the position angle in the regime where the opacity $\ga 1$.  This change, however, would be strongly correlated with a change in the polarization fraction.  The apparent
stability of the polarization fraction over 5 years with as much as 60 degrees
change in position angle argues against this hypothesis. 

On the other hand, a change of $\Delta {\rm RM}$ will
not lead to a change in the polarization fraction or
total intensity.  Both the magnitude of $\Delta {\rm RM}$ and the
timescale for its change are consistent with model expectations.

The RM as a function of radius from the black hole
can be calculated for different models
using a knowledge of the radial structure of the electron density, magnetic
field and electron temperature  along with an assumption of equipartition
between kinetic and magnetic energy densities
\citep{1999ApJ...527..851B,2000ApJ...545..842Q}.
For the case of CDAF models with an accretion rate
$\la 10^{-7} \msuny$, the electron density is not strongly peaked at the black hole
($\propto r^{-1/2}$)
and the RM $\la 3 \times 10^6 \rdm$ at all radii.
The actual radius at which the RM peaks is sensitive to the electron
temperature distribution.  For an electron temperature that
peaks at $3 \times 10^{11}$ K at a radius of $10 R_s$ and falls off 
inversely with radius, then the RM is greater than 
$3 \times 10^5 \rdm$ at radii $\ga 30 R_s$.  Thus, the change in
polarization angle could be due to a change in the electron density
and/or magnetic field at any radius of the accretion region
$\ga 30 R_s$.  On the other hand,
in the case of a steeply peaked electron density ($\propto
r^{-3/2}$) such as that required for the Bondi solution, the RM is dominated
by material very close to the black hole.  

The time scale of variability is only roughly determined by these
observations.  We see variability of the polarization angle on a time scale of 180
days followed by stability over 450 days.  We see no change in the polarization fraction
on time scales of hours, although our constraint is not very strong.
The predicted time scale of a turbulent change in the RM is comparable to the 
viscous time scale at the radius $r$ of the turbulence.  Given the range of
radii at which turbulent fluctuations could occur, we predict that RM fluctuations could occur on time scales of $10^{-1}$ to $10^3$ days.  Our current constraint on the time scale
of variability is too poor to determine at what radius the turbulence is taking
place.

Higher sensitivity mm $\lambda$ polarimetry obtained on scales from days to years by
the next generation millimeter observatories such as CARMA, the SMA
and ALMA will be able to generate a position angle and RM structure
function that can be matched to the detailed predictions of models.
Observations over a broader range of wavelengths are necessary
to clearly discriminate between intrinsic changes and RM changes.
Ultimately, these measurements can determine the mode of accretion
onto Sgr A*.

\section{Summary}

We have described new observations of linear polarization at 1.3 mm wavelength
from Sgr A*.   The polarization fraction is steady over several
years while the position angle changes by $\sim 30$ to 60 degrees on a time scale
of months.  The magnitude
and time scale of the position angle change are
consistent with the expectations of turbulence in the
accretion region surrounding Sgr A* at radii of 10 to 1000 $R_s$.  
This evidence supports the concept that most of the material 
that begins to accrete onto Sgr A* at the Bondi radius is lost in
a wind or outflow before it accretes onto the black hole itself, leaving
us with the very low luminosity source that we see.


\begin{deluxetable}{lrrrrrrr}
\tablecaption{Polarized and Total Flux Density of Sgr A* at 1.3 mm}
\tablehead{\colhead{Date} & \colhead{$\nu$} & \colhead{$I$} & \colhead{$Q$} & \colhead{$U$} & \colhead{$V$} & \colhead{$p$} & \colhead{$\chi$} \\
                    &\colhead{(GHz)} &  \colhead{(Jy)}& \colhead{(mJy)} &\colhead{(mJy)} &\colhead{(mJy)} & \colhead{(\%)} & \colhead{(deg)} }
\startdata
 14 OCT 2002 & 227.7 &  $   1.12 \pm   0.04$ & $   -69 \pm    39 $ & $   -57 \pm    39 $ & $    -3 \pm    39 $ & $   8.0 \pm   3.5 $ & $   110 \pm    13 $ \\ 
\dots & 230.5 &  $   1.17 \pm   0.04$ & $   -91 \pm    40 $ & $   -64 \pm    40 $ & $   -32 \pm    40 $ & $   9.5 \pm   3.4 $ & $   108 \pm    10 $ \\ 
\cline{2-8} 
\dots & 229.1 &  $   1.14 \pm   0.03$ & $   -80 \pm    28 $ & $   -60 \pm    28 $ & $   -17 \pm    28 $ & $   8.8 \pm   2.9 $ & $   109 \pm     8 $ \\ 
\\ 
 17 OCT 2002 & 215.0 &  $   0.69 \pm   0.03$ & $    -8 \pm    26 $ & $   -65 \pm    26 $ & $   -36 \pm    26 $ & $   9.5 \pm   3.8 $ & $   131 \pm    11 $ \\ 
\dots & 217.8 &  $   0.71 \pm   0.03$ & $   -74 \pm    26 $ & $   -24 \pm    26 $ & $   -47 \pm    26 $ & $  11.0 \pm   3.7 $ & $    99 \pm    10 $ \\ 
\cline{2-8} 
\dots & 216.4 &  $   0.70 \pm   0.02$ & $   -42 \pm    19 $ & $   -45 \pm    19 $ & $   -42 \pm    19 $ & $   8.7 \pm   3.3 $ & $   113 \pm     9 $ \\ 
\\ 
 19 MAY 2003 & 227.7 &  $   1.25 \pm   0.04$ & $  -180 \pm    37 $ & $   -96 \pm    37 $ & $   -22 \pm    37 $ & $  16.3 \pm   2.9 $ & $   104 \pm     5 $ \\ 
\dots & 230.5 &  $   1.20 \pm   0.04$ & $   -78 \pm    38 $ & $  -113 \pm    38 $ & $   -27 \pm    38 $ & $  11.5 \pm   3.2 $ & $   118 \pm     8 $ \\ 
\cline{2-8} 
\dots & 229.1 &  $   1.23 \pm   0.03$ & $  -131 \pm    26 $ & $  -104 \pm    26 $ & $   -24 \pm    26 $ & $  13.6 \pm   2.5 $ & $   109 \pm     4 $ \\ 
\\ 
 27 DEC 2003 & 227.7 &  $   0.89 \pm   0.03$ & $   -90 \pm    32 $ & $   -15 \pm    32 $ & $    -9 \pm    32 $ & $  10.2 \pm   3.6 $ & $    95 \pm    10 $ \\ 
\dots & 230.5 &  $   0.84 \pm   0.03$ & $   -43 \pm    34 $ & $    34 \pm    34 $ & $    16 \pm    34 $ & $   6.4 \pm   4.0 $ & $    71 \pm    18 $ \\ 
\cline{2-8} 
\dots & 229.1 &  $   0.87 \pm   0.02$ & $   -67 \pm    23 $ & $     9 \pm    23 $ & $     3 \pm    23 $ & $   7.8 \pm   2.7 $ & $    86 \pm    10 $ \\ 
\\ 
 05 JAN 2004 & 215.0 &  $   1.50 \pm   0.05$ & $   -92 \pm    50 $ & $  -136 \pm    50 $ & $   -82 \pm    50 $ & $  10.9 \pm   3.4 $ & $   118 \pm     9 $ \\ 
\dots & 217.8 &  $   1.53 \pm   0.05$ & $   -86 \pm    50 $ & $   -86 \pm    50 $ & $   -89 \pm    50 $ & $   7.9 \pm   3.3 $ & $   112 \pm    12 $ \\ 
\cline{2-8} 
\dots & 216.4 &  $   1.52 \pm   0.04$ & $   -89 \pm    36 $ & $  -111 \pm    36 $ & $   -85 \pm    36 $ & $   9.4 \pm   3.0 $ & $   116 \pm     7 $ \\ 
\\ 
\enddata
\end{deluxetable}

\plotone{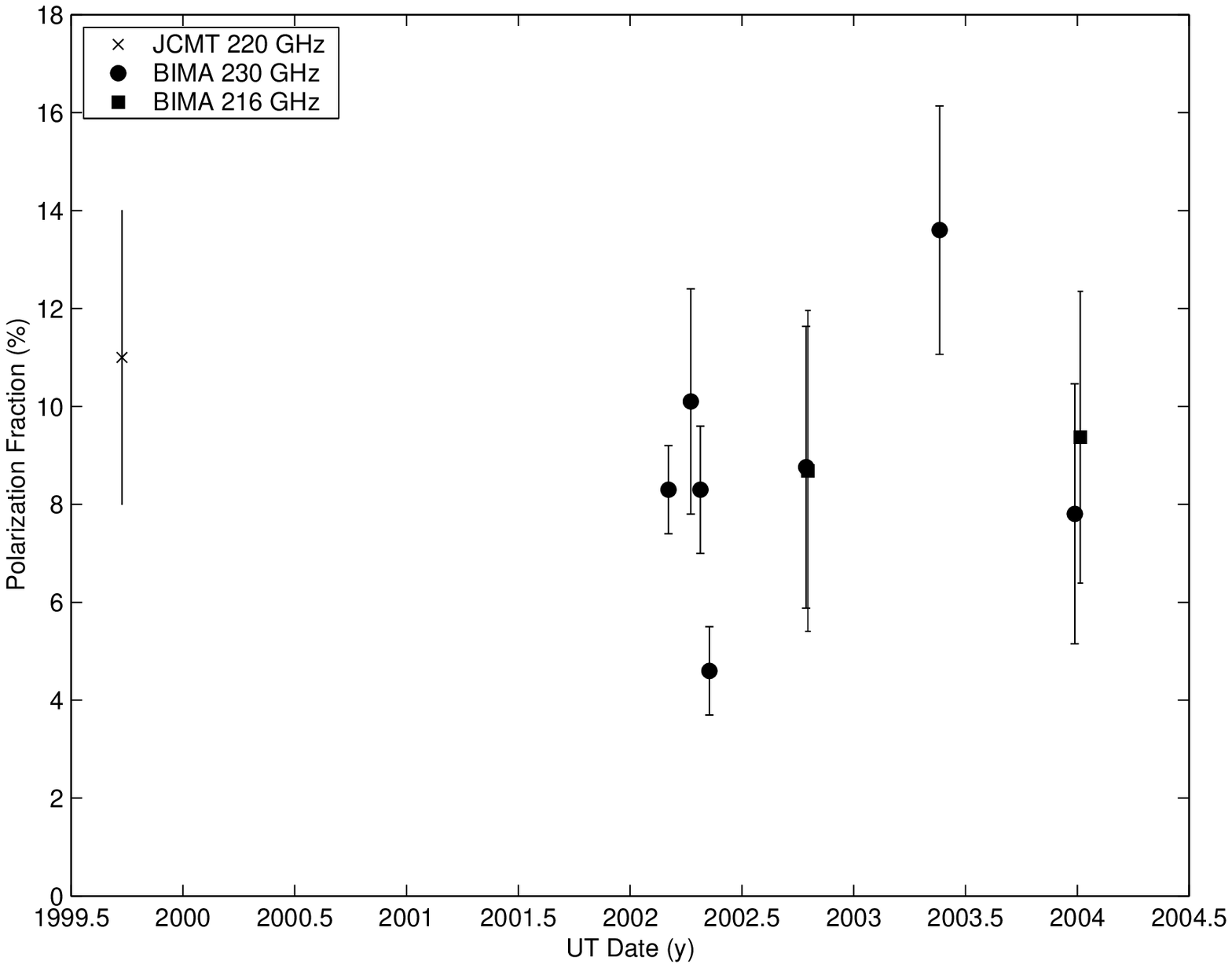}
\figcaption[pfrac.eps]{Fractional polarization at 1.3 mm as a function of time. We plot our
new results at 216 and 230 GHz, results from Bower \etal (2003) at 230 GHz
(between 2002 and 2002.5)
and results from Aitken \etal (2000) at 220 GHz.  \label{fig:pfrac}}

\plotone{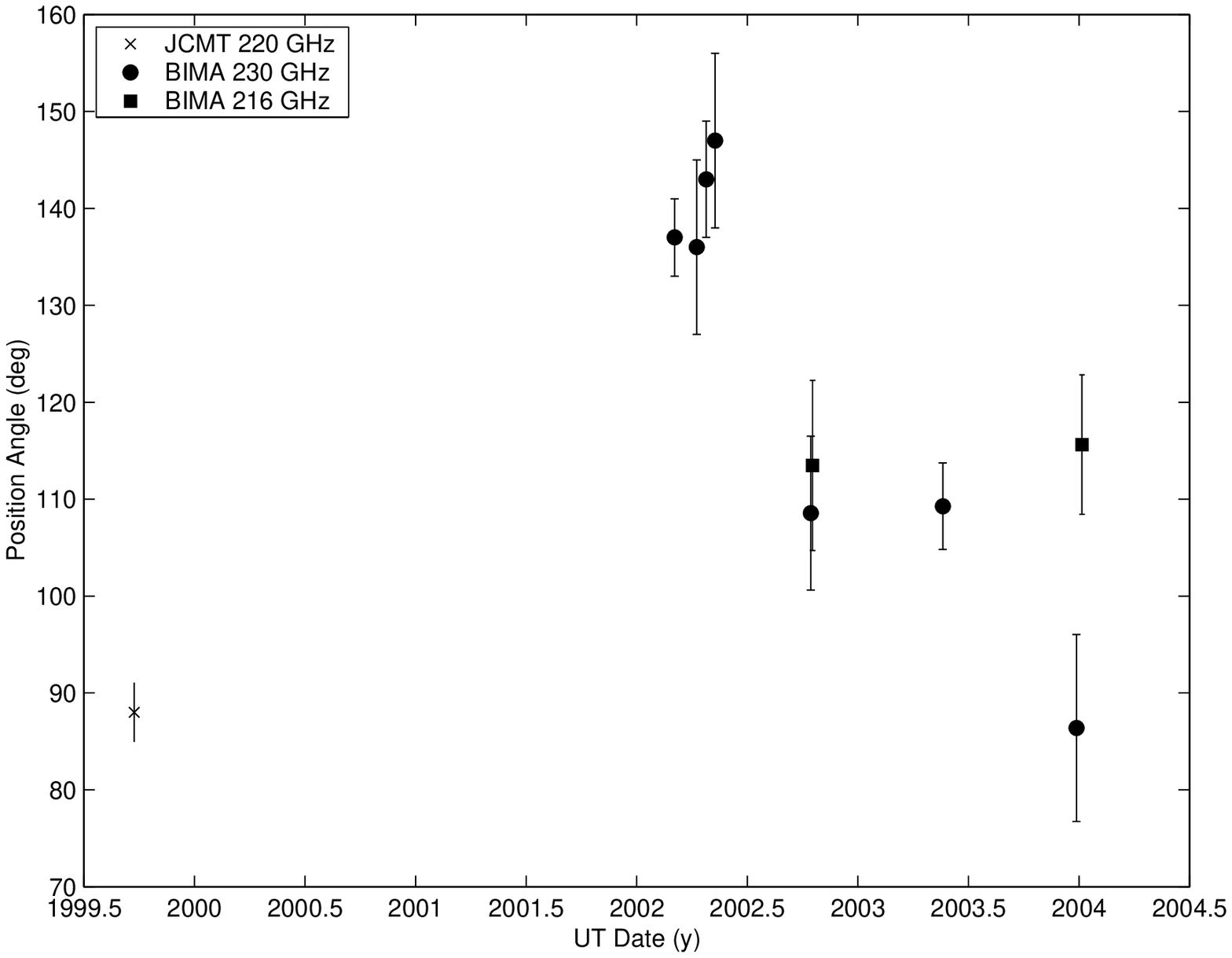}
\figcaption[pangle.eps]{Position angle of the linear polarization at 1.3 mm 
as a function of 
time. Symbols are the same as in Figure~\ref{fig:pfrac}. \label{fig:pangle}}

\end{document}